\documentclass[runningheads]{llncs}
\usepackage[english]{babel}

\usepackage{array} 
\usepackage{graphicx}
\usepackage{amsmath}
\usepackage{graphicx}
\usepackage[colorlinks=true, allcolors=blue]{hyperref}
\usepackage{subcaption}
\usepackage{enumitem} 
\usepackage{adjustbox}
\usepackage{times}
\usepackage{multirow}
\usepackage{tikzpagenodes}
\newcommand{\roundedref}[1]{\tikz[baseline=(char.base)] \node[draw, rounded corners=2pt, inner sep=3pt] (char) {#1};}

\title{OCPM$^2$: Extending the Process Mining Methodology for Object-Centric Event Data Extraction}
\titlerunning{OCPM$^2$: Extending the PM$^2$ for OCED Extraction}

\author{Najmeh Miri\inst{1}
\and Shahrzad Khayatbashi\inst{2}
\and Jelena Zdravkovic\inst{1}
\and Amin Jalali\inst{1}
 }

\authorrunning{Najmeh Miri et al.}

\institute{Stockholm University, Stockholm, Sweden,\\
\email{(najmeh.miri | jelenaz | aj)@dsv.su.se}
\and
    Linköping University, Linköping, Sweden,\\
    \email{shahrzad.khayatbashi@liu.se}
}

\begin{document}
\maketitle

\begin{abstract}

Object-Centric Process Mining (OCPM) enables business process analysis from multiple perspectives. For example, an educational path can be examined from the viewpoints of students, teachers, and groups. This analysis depends on Object-Centric Event Data (OCED), which captures relationships between events and object types, representing different perspectives. Unlike traditional process mining techniques, extracting OCED minimizes the need for repeated log extractions when shifting the analytical focus. However, recording these complex relationships increases the complexity of the log extraction process.
To address this challenge, this paper proposes a methodology for extracting OCED based on PM\inst{2}, a well-established process mining framework. Our approach introduces a structured framework that guides data analysts and engineers in extracting OCED for process analysis. We validate this framework by applying it in a real-world educational setting, demonstrating its effectiveness in extracting an Object-Centric Event Log (OCEL), which serves as the standard format for recording OCED, from a learning management system and an administrative grading system.
\vspace{-0.7\baselineskip}	
\keywords {Object-Centric Process Mining, Methodology, Log Extraction}

\end{abstract}

\vspace{-2.1\baselineskip}	

\section{Introduction}\label{sec:Introduction}
\vspace{-0.7\baselineskip}	
Object-Centric Process Mining (OCPM) is a data-driven approach that enables multi-perspective process analysis, allowing for a more comprehensive understanding of complex workflows~\cite{ocel2023specification,van2019object}. Unlike traditional process mining techniques that focus on a single perspective - such as analyzing the learning process from the student perspective alone - OCPM considers multiple object types simultaneously, revealing hidden dependencies and bottlenecks that might remain undetected when examining processes from a singular viewpoint~\cite{2024ExploringObjectCentric}. For example, in an educational setting, the learning process can be analyzed not only from the perspective of students but also from that of teachers and course groups, enabling insights into instructor-student interactions, group-based collaboration, and administrative workflows.

To apply OCPM in practice, Object-Centric Event Data (OCED)~\cite{fahland2024towards} must be extracted to capture events in relation to multiple object types. In an educational setting, typical object types include students, teachers, and groups, each of which plays a distinct role in the learning process. Unlike traditional event logs, OCED reduces the need for repeated log extraction when shifting between analytical perspectives. Several formats exist for representing OCED, including the Object-Centric Event Log (OCEL)~\cite{ocel2023specification}, Data-Aware Object-Centric Event Logs (DOCEL)~\cite{goossens2022enhancing}, Event Knowledge Graph (EKG)~\cite{esser2021multi}, and Temporal Event Knowledge Graph (tEKG)~\cite{khayatbashi2024transforming}, which can be transformed into one another~\cite{khayatbashi2023transforming,khayatbashi2024transforming}. However, extracting OCED from operational systems remains inherently complex, as it requires capturing relationships between events and all possible object types while ensuring proper data integration.

Despite recent advancements, OCED extraction from information systems continues to pose significant challenges, necessitating new methods and guidelines to support researchers and practitioners. While prior studies~\cite{2023SAPExtractionCaseStudy,2022SAPExtractionApproach,2023VKG} have addressed specific challenges in OCED extraction, existing solutions are often system- or platform- dependent, tailored to particular environments such as OnProm~\cite{2023VKG} or designed for extracting logs from Enterprise Resource Planning (ERP) systems~\cite{2023SAPExtractionCaseStudy,2022SAPExtractionApproach}.

To bridge this gap, this study proposes a structured and system-agnostic framework for OCED extraction, addressing the complexities associated with capturing multi-perspective event data. Our approach extends PM\inst{2}~\cite{2015PM2}, a well-established process mining framework, offering a systematic framework to support data analysts and engineers in extracting event logs for process analysis. To demonstrate its practical applicability, we applied our framework in the educational domain, extracting OCEL from a learning management system and an administrative grading system. This enables the simultaneous analysis of educational paths from both the group and student perspectives, allowing for a deeper understanding of learning dynamics and collaborative processes.

The remainder of this paper is structured as follows.
Section~\ref{sec:Background} elaborates on previous process mining methodologies, provides an overview of two primary OCED formats, and summarizes existing object-centric event data extraction approaches.
Section~\ref{sec:Method} presents our OCED extraction method.
Section~\ref{sec:CseStudy} details the application of our method for extracting OCEL from learning management systems.
Finally, Section~\ref{sec:Conclusion} concludes the paper.

\vspace{-0.7\baselineskip}	
\section{Background}\label{sec:Background}
\vspace{-0.7\baselineskip}	
This section provides an overview of existing process mining methodologies and peer-reviewed approaches for object-centric event data (OCED) extraction. In addition, it explores two primary approaches for representing OCED.

\vspace{-1\baselineskip}	
\subsection{Process Mining Methodologies}
\vspace{-0.7\baselineskip}	
Process Diagnostics Method (PDM)~\cite{2009PDM-Bozkaya}, the L* Life-Cycle Model~\cite{der2011discovery}, and PM\textsuperscript{2}~\cite{2015PM2} are three well-known process mining methodologies followed by different people to apply process mining in practice. 
The Process Diagnostics Method (PDM)~\cite{2009PDM-Bozkaya} focuses on rapid, high-level analysis of business processes without requiring domain-specific knowledge. It examines control flow, performance, and organizational aspects based on event logs, providing quick insights. However, PDM lacks explicit iterative refinement of the analysis~\cite{2015PM2}, and is less suitable for long-term process optimization efforts~\cite{2009PDM-Bozkaya}.

In contrast, the L* Life-Cycle Model~\cite{der2011discovery} follows a structured, stepwise approach that includes planning and justification, event log extraction, process model discovery, integration, and operational support. It defines three distinct project approaches: goal-driven (focusing on achieving predefined objectives), question-driven (addressing specific business queries), and data-driven (exploring available data for insights). However, this methodology is effective solely for structured processes where a single integrated process model can be developed.

To overcome the limitations of both PDM and L*, PM\textsuperscript{2}~\cite{2015PM2} introduces an iterative process mining methodology that incorporates stakeholder involvement and a structured roadmap for continuous process improvement. Unlike L*, which primarily targets structured processes, PM\textsuperscript{2} accommodates both structured and unstructured processes, allowing for evolving business-related questions and iterative refinements. 

The PM\textsuperscript{2} methodology consists of six key \textit{stages}~\cite{2015PM2}: (i) \textit{Planning}, which involves defining project goals, such as performance improvement or compliance verification, formulating research questions (we refer to as business-related questions throughout the paper), and identifying relevant business processes. This stage also involves assembling a multidisciplinary
project team, including business experts, system specialists, process analysts, and business owners; (ii) \textit{Extraction}, where event data is collected from information systems, considering data quality, scope, and process knowledge transfer; (iii) \textit{Data Processing}, which prepares event logs by defining case notions, filtering, aggregating, and enriching event data; (iv) \textit{Mining and Analysis}, where process discovery, conformance checking, and performance enhancement techniques are applied; (v) \textit{Evaluation}, which includes interpreting analysis results, validating insights with domain experts, and refining business-related questions; and (vi) \textit{Process Improvement and Support}, where process changes are implemented, or operational support is provided based on the findings.

These methodologies offer structured approaches for applying process mining in practice and are well-suited for extracting traditional event logs from source systems. However, extracting OCED presents additional complexities due to the involvement of multiple object types, each requiring the capture of different events. As a result, analysts can easily overlook some necessary relationships, leading to repeated data extraction. This redundancy creates bottlenecks in the log extraction process. To address this challenge, we extend PM$^2$ to enhance the efficiency of OCED extraction. Before introducing our approach in the next section, the following subsections provide an overview of OCED formats and discuss existing methodologies for OCED extraction.

\vspace{-0.7\baselineskip}	
\subsection{Brief Overview of OCED}
\vspace{-0.7\baselineskip}	

\begin{figure}[t!]
    \centering
    \begin{adjustbox}{width=1\textwidth,center}
        \begin{subfigure}{0.5\linewidth}
            \centering
            \includegraphics[width=\linewidth]{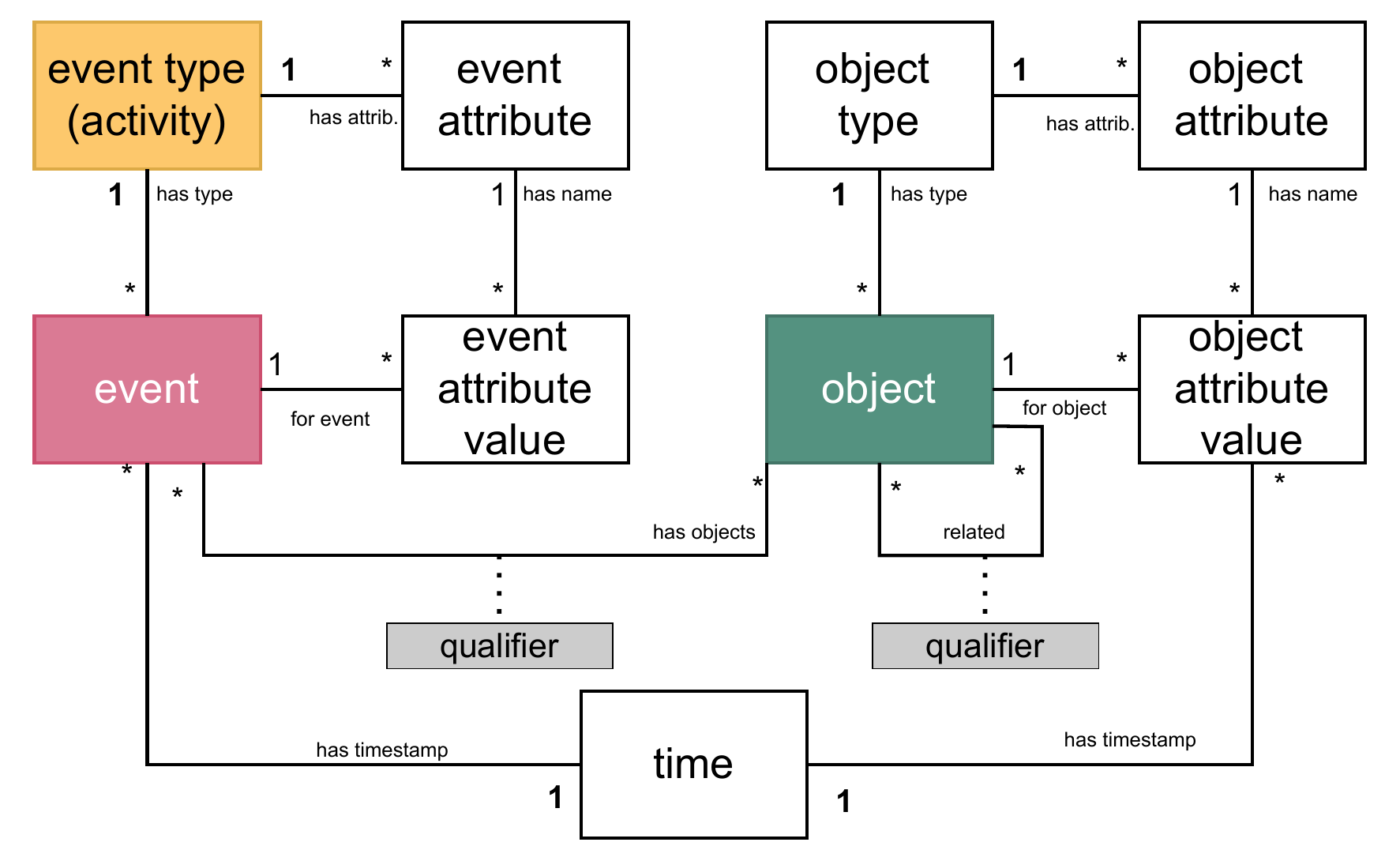}
            \caption{OCEL 2.0 metamodel~\cite{ocel2023specification}}
            \label{fig:first}
        \end{subfigure}
        \begin{subfigure}{0.5\linewidth}
            \centering
            \includegraphics[width=\linewidth]{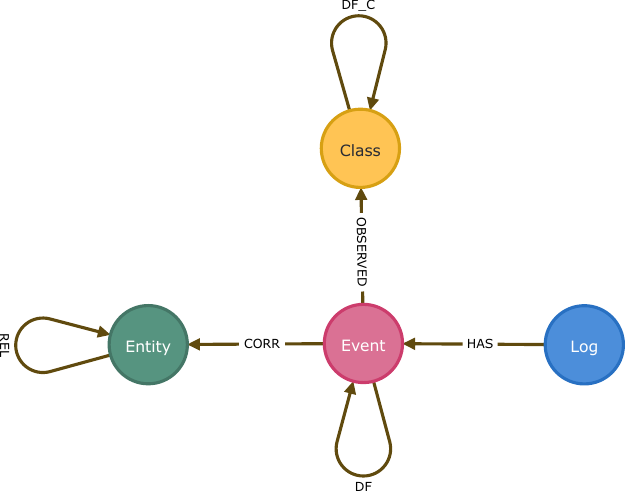}
            \caption{EKG metamodel~\cite{esser2021multi}}
            \label{fig:second}
        \end{subfigure}
    \end{adjustbox}
    \caption{Two metamodels for storing OCED}
    \label{fig:metamodels}
    \vspace{-1.5\baselineskip}	
\end{figure}

Object-Centric Event Data (OCED) provides a richer, more realistic representation of process event data by capturing multiple objects, their relationships, and interactions. Unlike traditional event logs, which focus on a single case notion (e.g., an order or a customer), OCED captures the relationships between different entities related to events.
\figurename~\ref{fig:metamodels} presents two metamodels for storing OCED: Object-Centric Event Logs (OCEL 2.0)~\cite{ocel2023specification} and Event Knowledge Graphs (EKG)~\cite{esser2021multi}.

The OCEL 2.0 metamodel, shown on the left side of \figurename~\ref{fig:metamodels}, defines an \textit{event} as an atomic action that occurs in a business process, representing the execution of an \textit{event type} (a.k.a, an \textit{activity}), such as ``place order" in Order-to-Cash process.
Each event type can have multiple \textit{event attributes}, and each event can store different \textit{event attribute values}. OCEL records \textit{objects} as separate entities, where each object belongs to an \textit{object type}. For example, the ``order" can be an object type with multiple instances representing different orders. Each object type can have multiple object attributes, and each object can store different values for these attributes at specific points in time. Objects can also be related to other objects through \textit{object-to-object} relations; for example, an order can contain multiple items. As illustrated in the metamodel, events can be associated with multiple objects and vice versa through \textit{event-to-object} relations.

EKG, on the right side of the figure, provides an alternative approach for capturing OCED by representing events, entities, logs, and classes as nodes with specific relationships among them. Events and classes in EKG correspond to events and event types in OCEL, respectively. Entities in EKG represent objects, with object types stored as a property of the entity node called \textit{EntityType}. Object-to-object relations are captured as edges between entity nodes labeled \textit{REL}, while event-to-object relations are represented as edges between event and entity nodes labeled \textit{CORR}.

There are more variations capturing OCED data like temporal EKG~\cite{khayatbashi2024transforming}, which improves EKGs by tracking how object attributes change over time so events can be analyzed using the correct object information at each point in time, and Data-aware Object-Centric Event Logs (DOCEL)~\cite{goossens2022enhancing}, which extends OCEL by tracking attribute changes over time, linking attributes to objects and events, and allowing attributes to have multiple values. 

\vspace{-0.7\baselineskip}	
\subsection{OCED Extraction Initiatives}
\vspace{-0.7\baselineskip}	

Several approaches have been developed to extract OCED from information systems, particularly Enterprise Resource Planning (ERP) systems. Here, we exclude methods that focus solely on transforming OCED formats from other formats or traditional event logs since this study focuses on proposing a method for OCED extraction directly from information systems.

Berti et al. proposed a method for extracting event data from SAP ERP systems using OCEL~\cite{2022SAPExtractionApproach}. Their approach begins by constructing a Graph of Relations (GoR) to map relevant business process tables. Once these tables are mapped, data are extracted and stored in the OCEL format. To streamline this process, they developed a Python-based tool that identifies relevant tables and their relationships. The extracted data can subsequently be used for OCPM or converted into traditional event logs.

Building on this approach, Berti et al. conducted a case study focusing on the Purchase-to-Pay (P2P) process~\cite{2023SAPExtractionCaseStudy}. Similar to their previous work~\cite{2022SAPExtractionApproach}, this study employed graph-based techniques to model relationships within SAP's complex data structures. The methodology utilized an object interaction graph to visualize object relationships and applied the PM\textsuperscript{2} framework to structure the extraction and analysis process within a well-defined methodological context.

Beyond SAP-specific solutions, Xiong et al. introduced a method for extracting OCEL from relational databases using a Virtual Knowledge Graph (VKG) approach~\cite{2023VKG}. Their method extended the OnProm framework to support both OCEL and the traditional XES format. By leveraging ontology-based data access (OBDA) and the VKG system Ontop, they defined domain ontologies and mappings, enabling event log extraction via SPARQL queries. Their approach was validated using the Dolibarr ERP system, demonstrating how relational data can be transformed into structured OCELs.

Unlike these system-specific approaches, this paper presents a generalized framework for OCED extraction that is adaptable across various information sources. Our methodology ensures broader applicability and ease of implementation across different systems by providing a step-by-step process that does not rely on platform-dependent techniques.

\vspace{-0.7\baselineskip}	
\section{The OCPM$^2$ Methodology}\label{sec:Method}
\vspace{-0.7\baselineskip}	
This section introduces OCPM$^2$, an Object-Centric Process Mining Methodology defined as an extension of PM\textsuperscript{2}~\cite{2015PM2} to facilitate the extraction of Object-Centric Event Data (OCED)~\cite{fahland2024towards}.
We define our method by extending PM$^2$ to ensure the reuse of a well-established and previously tested methodology for event log extraction, originally designed for traditional process mining. Our approach retains the clarity and efficiency of PM$^2$ while making it suitable for extracting object-centric event data from diverse information systems. We also presented our methodology to 12 domain experts in a workshop, which helped us further confirm the OCPM$^2$ validity.

\begin{figure}[t!]
    \adjustbox{center}{\includegraphics[width=1.1\linewidth]{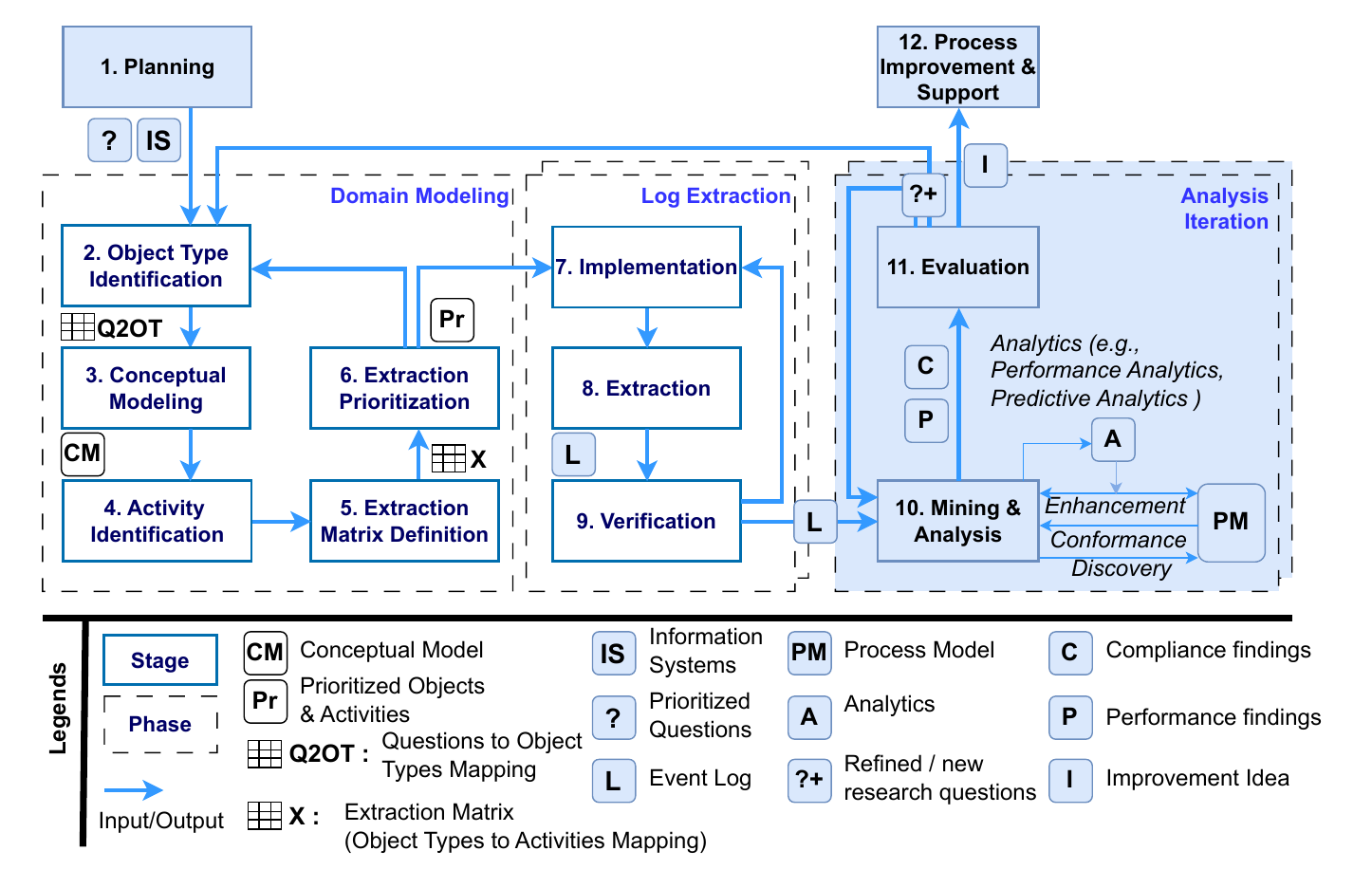}}
    
    \caption{An overview of the OCPM$^2$ methodology extending the PM$^2$ methodology~\cite{2015PM2}}
    \label{fig:Method_Representation}
	\vspace{-1.5\baselineskip}	
\end{figure}

The methodology consists of twelve stages, where the related stages are grouped into phases, as indicated by dashed rectangles in \figurename~\ref{fig:Method_Representation}. The process begins with the \textit{Planning} stage, followed by the \textit{Domain Modeling}, \textit{Log Extraction}, and \textit{Analysis Iteration} phases, and concludes with the \textit{Process Improvement \& Support} stage. Each stage produces specific artifacts, as outlined in the figure. The following sections elaborate on these stages, the generated artifacts, and the defined phases. The filled blue colored phases and stages are identical to those in the PM\textsuperscript{2} methodology briefly described in Section~\ref{sec:Background} (we refer readers to the PM\textsuperscript{2}~\cite{2015PM2} for further details).

\vspace{-0.7\baselineskip}	
\subsection{Planning}
The \textit{Planning} stage follows PM\textsuperscript{2}, producing the following key artifacts: i) prioritized business-related questions (denoted by \roundedref{\texttt{?}} in the \figurename~\ref{fig:Method_Representation}), ii) a list of information systems (\roundedref{\texttt{IS}}) from which data will be extracted to answer these questions.
The minor difference from PM\textsuperscript{2} is that the questions are prioritized, allowing for prioritizing the object type extraction later.

\vspace{-0.7\baselineskip}	
\subsection{Domain Modeling}\label{subsec:domain}
The \textit{Domain Modeling} phase consists of five stages (stages 2 to 6 in \figurename~\ref{fig:Method_Representation}), identifying and conceptualizing the information needed for the extraction of OCED.
As OCED involves multiple object types, a domain model can act as a foundation to facilitate identifying the involved object types, their relations, activities, and the relations between object types and events. This information can be gathered through interviews with business experts, reviewing system documentation, and analyzing the underlying database schema. Without this foundation, the extraction process may result in an incomplete OCED.

The phase begins with \textit{Object Type Identification} to determine all object types needed to address the questions elicited in the \textit{Planing}. Because an object type may be required for multiple questions, we propose documenting this using a Question-to-Object Type Matrix (\texttt{Q2OT}), inspired from the bus matrix which is a widely used technqiue in identifying dimensions and related processes in data warehousing~\cite{kimball2008data}. This matrix represents questions as rows and object types as columns, marking intersection cells to indicate the required object types for each question.

Next, in the \textit{Conceptual Modeling} stage, process analysts and business experts collaborate to define a model (\roundedref{\texttt{CM}}) that captures the relationships between object types. Business experts provide domain knowledge, while process analysts structure this information using a conceptual modeling language to represent key relationships. Among these, recognizing the ``is-a" relationship—where one object type is a specialized version of another (e.g., Student is a User)— enables capturing OLAP operations such as drill-down and roll-up in OCPM based on OCEL~\cite{2025uncovering_patterns,khayatbashi2024advancing}.

The \textit{Activity Identification} stage follows, eliciting different activities (or event types) to which the identified object types and questions are related. These activities can be identified in various ways, such as analyzing activities directly related to the formulated questions, reviewing system documentation, and consulting business experts.

The \textit{Extraction Matrix Definition} stage defines the extraction matrix (\texttt{X}), where the relationships between object types and activities are documented. When an ``is-a" relationship exists among object types, separate columns can specify whether an activity relates to the supertype class or a subtype class. 
Various documentation approaches can be adapted based on specific perspectives. From a system perspective, CRUD (Create, Read, Update, and Delete) operations are suitable for specifying the type of relationship between an activity and an object type. However, from a process perspective, CRUD may not fully capture the dynamic nature of object interactions. For example, when a user views a module such as File, CRUD cannot document the role of user properly. Instead, documenting the quantity range of objects for each object type provides a more process-oriented approach. This method captures the expected number of objects per event and supports data validation by ensuring that the extracted log adheres to these expectations. The details of these adaptations can be delegated to design and development teams.

Finally, in the \textit{Extraction Prioritization} stage, business, process, and system experts collaborate to prioritize the extraction of objects and events based on their i) feasibility, ii) data availability, and iii) importance. Given the potentially large number of object types and activities, extracting all data at once may not be feasible due to data being fragmented across multiple sources and the need to check data quality step-by-step to catch errors early. To manage this, we propose a prioritization artifact as the output of this stage, including a set of prioritized objects and activities (\roundedref{\texttt{Pr}}).
Note that new object types can be identified during this phase, requiring the phase to be repeated to have a more holistic understanding of the domain. 

\vspace{-0.7\baselineskip}	
\subsection{Log Extraction}

The \textit{Log Extraction} phase in OCPM$^2$ consists of three stages (Stages 7 to 9 in \figurename~\ref{fig:Method_Representation}) as follows:

In the \textit{Implementation} stage, developers build pipelines to extract data from source systems, transform them into the desired OCED format, and load them into destination storage.
If OCEL is the target format, objects should be processed first based on their list in the extraction matrix. When an ``is-a" relationship exists among object types, we propose storing objects at the supertype level with an attribute distinguishing subtypes. This structure follows the single-table inheritance technique used in Object-Relational Mapping (ORM)~\cite{fowler2012patterns}. This ensures referential integrity and facilitates efficient querying and aggregation, allowing drill-down operations~\cite{khayatbashi2024advancing,khayatbashi2025olap}.
Subsequently, developers implement extraction logic for object-to-object relationships, determined by the conceptual model (stage 3), in parallel with implementing extraction logic for events based on the extraction matrix. Once event extraction is complete, they implement extraction logic for event-to-object relationships.
Please note that similar mapping can be done for EKG~\cite{esser2021multi} and tEKG~\cite{khayatbashi2023transforming} as the OCED formats are transformable to each other.

The \textit{Extraction} stage is separated from \textit{Implementation} because it enables the development of data pipelines to extract, clean, and integrate data for process mining as well as testing the result before final extraction in a production environment.

In the \textit{Verification} stage, correctness can automatically be checked by deriving a matrix from the extracted event data. 
This matrix shows the number of object types for each event type, ensuring that documented relationships in the Extraction Matrix are properly implemented.

Although the log extraction process has been adapted to extract OCED, the remaining stages remain identical to PM\textsuperscript{2}~\cite{2015PM2} since the types of applicable process mining techniques are not based on analyzing case-centric or object-centric event logs. However, these phases may include new process mining techniques in the future as the area develops further. Additionally, one can reduce the dimensionality of OCED by converting the log to the traditional format by correlating events to only one object type, a.k.a. flattening, to enable the use of traditional process discovery techniques. The flattening can also help filter the log by identifying similar object type clusters in OCEL~\cite{jalali2022object}, meaning that process mining techniques can be applied to a range of case-based and object-centric event logs. 

\vspace{-0.7\baselineskip}	
\subsection{Analysis Iteration}
The \textit{Analysis Iteration} phase in PM\textsuperscript{2} supports gaining insights from extracted event data through two stages: \textit{Mining \& Analysis} and \textit{Evaluation}. 
The \textit{Mining \& Analysis} stage includes process model discovery (\roundedref{\texttt{PM}}) from the input log (\roundedref{\texttt{L}}), conformance checking, and process enhancement based on analytical insights (\roundedref{\texttt{A}}) resulting from various analytical techniques, such as performance, and predictive analysis~\cite{vanderAalst2022}. This can yield compliance findings (\roundedref{\texttt{C}}) and performance findings (\roundedref{\texttt{P}}).
The \textit{Evaluation} stage results in process improvement ideas (\roundedref{\texttt{I}}) or refined/new process-related questions (\roundedref{\texttt{?+}}), which may prompt further analysis or new data extraction.

\vspace{-0.7\baselineskip}	
\subsection{Process Improvement \& Support}
The \textit{Process Improvement \& Support} stage applies the insights from process analysis to optimize workflows and enhance performance. It involves defining strategies, supporting implementation, and monitoring changes.

\vspace{-0.7\baselineskip}	
\section{Case Study}\label{sec:CseStudy}
\vspace{-0.7\baselineskip}	
This section presents the application of the OCPM$^2$ methodology in a process mining project conducted at the Department of Computer and Systems Sciences, Stockholm University. We elaborate on each stage and phase below.
In addition, the second and last authors conducted a separate feasibility study demonstrating the applicability of this methodology in a real-world case from the insurance domain~\cite{khayabashi2025if}.

\vspace{-0.7\baselineskip}	
\subsection{Planning}
\vspace{-0.7\baselineskip}	
This project was initiated to explore how process mining can provide teachers with data-driven insights crucial for improving their educational content and tasks' efficiency. To determine the most important questions and their relative significance, we conducted interviews with 10 teachers. The interviews were recorded, transcribed, anonymized, and analyzed to extract key questions and assess their perceived importance.
The outcome of this stage consisted of two artifacts: a set of prioritized questions compiled from the interviews (\roundedref{\texttt{?}}) and a list of relevant information systems (\roundedref{\texttt{IS}}).
The four most important questions identified were:
\begin{itemize}[leftmargin=*]
    \item Q1: What learning paths do students typically follow when accessing educational materials, such as files, pages, and folders, throughout a course?
    \item Q2: How do students submit (or resubmit) individual and group assignments during the course?
    \item Q3: Do students who usually take the lead in submitting group assignments on behalf of the group tend to achieve higher final grades (exam and assignment grades)?
    \item Q4: How is students' exam success related to their frequency of accessing course materials, such as files, pages, and folders?
\end{itemize}

Students often follow different routes when accessing educational materials, influencing their engagement and performance. Thus, analyzing students' learning paths (Q1) provides insights into how these behaviors influence the outcomes while also identifying the challenging areas that affect their success~\cite{2025AnEduPMModel}. Examining assignment submission patterns (Q2) helps to assess student engagement and identify potential challenges in meeting deadlines. Furthermore, analyzing whether high-achieving students demonstrate different behaviors, such as taking the lead in submitting group assignments (Q3), provides insights into collaborative learning strategies, such as encouraging role rotation. Finally, analyzing the relationship between course material access and exam performance (Q4) can help educators identify patterns that distinguish successful students from those who struggle. 

Although we identified numerous questions to guide the project, we present only a subset here to illustrate the application of the methodology. The information systems identified at this stage included 1) the learning management system, Moodle~\cite{2024moodle}, and 2) the platform that records students' grades. 

\subsection{Domain Modeling}

\textbf{Object Type Identification:}
From the questions above, we identified the following object types: \textit{Student}, \textit{File}, \textit{Page}, \textit{Folder}, \textit{Assignment}, \textit{Group}, \textit{Course}, and \textit{Exam}. These object types were identified by extracting key nouns from the questions that belong to the domain vocabulary.
These object types model the domain of interest.
\tablename~\ref{tab:Q2OT} presents the Question-to-Object Type matrix (\texttt{Q2OT}), derived from the previous stage.

\vspace{-1\baselineskip}	
\begin{table}[h!]
    \centering
    \begin{tabular}{l|c|c|c|c|c|c|c|c}
       & \multicolumn{1}{l|}{Student} & \multicolumn{1}{l|}{File} & \multicolumn{1}{l|}{Page} & \multicolumn{1}{l|}{Folder} & \multicolumn{1}{l|}{Assignment} & \multicolumn{1}{l|}{Group} & \multicolumn{1}{l|}{Course} & \multicolumn{1}{l}{Exam} \\ \hline
    Q1 & *                  & *             & *                         & *                         &                                 &                            & *                           &                          \\
    Q2 & *                 &                 &                           &                           & *                               & *                          & *                           &                         \\
    Q3 & *                       &           &                           &                           & *                               & *                          & *                           & *                        \\
    Q4 & *               & *                & *                         & *                         &                                 &                            & *                           & *                       
    \end{tabular}
    \vspace{10pt} 
    \caption{\centering Question-to-Object Type matrix supporting object type identification}
    \label{tab:Q2OT}
    \vspace{-2.5\baselineskip}	
\end{table}

\begin{figure}[t!]
    \adjustbox{center}{\includegraphics[width=1.1\linewidth]{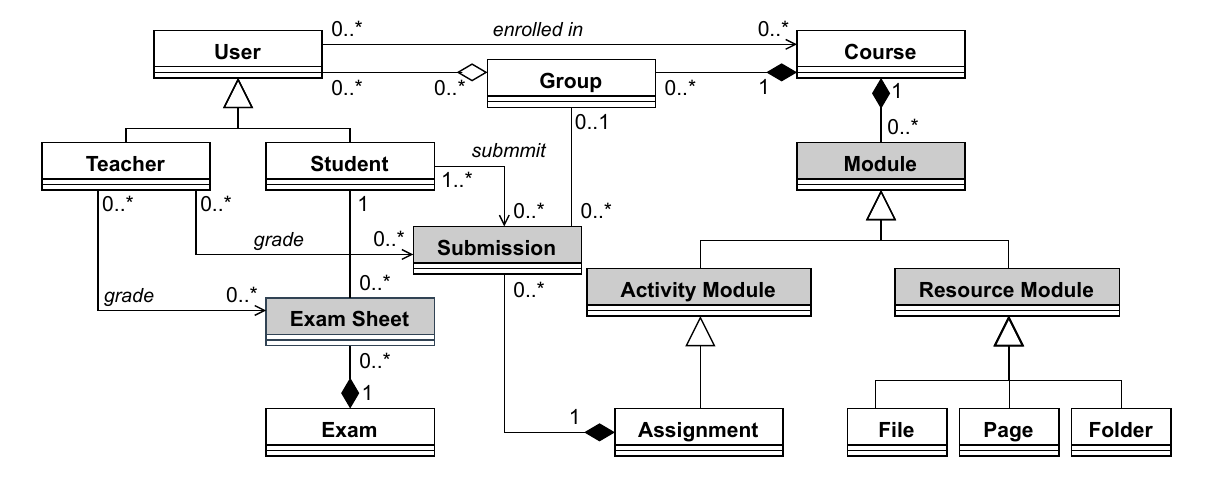}}
    \caption{The conceptual model illustrating object type relationships}
    \label{fig:DomainModel}
    \vspace{-1.5\baselineskip}	
\end{figure}

\noindent
\textbf{Conceptual Modeling:}
\figurename~\ref{fig:DomainModel} presents the conceptual model using UML notation, a de facto modeling standard widely used in the industry, illustrating the relationships between object types. Nonetheless, the methodology is modeling-language agnostic, and equivalent results can be achieved with alternative notations. This model was developed following the conceptual modeling process described in Section~\ref{subsec:domain}, based on stakeholder input, system documentation, and database schema analysis. The model includes additional object types beyond those represented in the \texttt{Q2OT} matrix, reflecting the iterative nature of our approach. For example, the object type \textit{Teacher} was later identified by reviewing the Moodle documentation as a relevant entity within the process. Additionally, \textit{User} was introduced as a supertype encompassing both \textit{Teacher} and \textit{Student} as explained in Section \ref{fig:Method_Representation}.

Design choices also influence object type identification. For instance, \textit{Grade} was modeled as an attribute of both \textit{Submission} and \textit{Exam Sheet}, though an alternative approach could define it as a separate class.
The model also includes grayed-out classes, such as \textit{Exam Sheet}, \textit{Submission}, \textit{Module}, \textit{Activity Module}, and \textit{Resource Module}. While some of these could be considered object types, their selection depends on modeling design choices.
For example, \textit{Exam} and \textit{Assignment} are tightly related to \textit{Exam Sheet} and \textit{Submission}, respectively, due to the decomposition relation nature. Thus, we have not considered defining separate object types for \textit{Exam Sheet} and \textit{Submission}, which are parts of \textit{Exam} and \textit{Assignment}. 

\vspace{0.5\baselineskip}	
\noindent
\textbf{Activity Identification:}
Activities were directly identifiable from the questions by examining the leading tasks (verbs) in the questions. 
For example, \textit{view file}, \textit{view page}, and \textit{view folder} were relevant for Q1, while \textit{submit assignment} and \textit{resubmit assignment} addressed Q2. Similarly, \textit{set assignment grade}, \textit{set exam grade}, and \textit{submit assignment} were relevant for Q3.
In addition, \textit{set exam grade}, \textit{view file}, \textit{view folder}, and \textit{view page} were relevant for Q4.

\vspace{0.5\baselineskip}	
\noindent
\textbf{Extraction Matrix Definition:}
\tablename~\ref{tab:Xmatrix} documents relationships between identified activities and object types. In this version, we recorded the possible number of instances for each object type within the table’s cells. To provide more detailed documentation, we included \textit{User}, the supertype of \textit{Teacher} and \textit{Student}, as a separate column.

\begin{table}[t!]
    \centering
    \begin{tabular}{l|ccc|c|c|c|c|c|c|c}
                         & \multicolumn{3}{c|}{User}                                          & \multirow{2}{*}{Exam} & \multirow{2}{*}{File} & \multirow{2}{*}{Page} & \multirow{2}{*}{Folder} & \multirow{2}{*}{Assignment} & \multirow{2}{*}{Group} & \multirow{2}{*}{Course} \\ \cline{2-4}
                         & \multicolumn{1}{c|}{User} & \multicolumn{1}{c|}{Teacher} & Student &                       &                       &                       &                         &                             &                        &                         \\ \hline
    view file            & \multicolumn{1}{c|}{1}    & \multicolumn{1}{c|}{}        &         &                       & 1                     &                       &                         &                             &                        & 1                       \\
    view page            & \multicolumn{1}{c|}{1}    & \multicolumn{1}{c|}{}        &         &                       &                       & 1                     &                         &                             &                        & 1                       \\
    view folder          & \multicolumn{1}{c|}{1}    & \multicolumn{1}{c|}{}        &         &                       &                       &                       & 1                       &                             &                        & 1                       \\
    submit assignment    & \multicolumn{1}{c|}{}     & \multicolumn{1}{c|}{}        & 1       &                       & 0..*                  &                       &                         & 1                           & 0..1                   & 1                       \\
    resubmit assignment  & \multicolumn{1}{c|}{}     & \multicolumn{1}{c|}{}        & 1       &                       & 0..*                  &                       &                         & 1                           & 0..1                   & 1                       \\
    set assignment grade & \multicolumn{1}{c|}{}     & \multicolumn{1}{c|}{1}       & 1       &                       & 0..*                  &                       &                         & 1                           & 0..1                   & 1                       \\
    set exam grade       & \multicolumn{1}{c|}{}     & \multicolumn{1}{c|}{1}       & 1..*    & 1                     &                       &                       &                         &                             &                        & 1                      
    \end{tabular}
    \vspace{10pt} 
    \caption{\centering The extraction matrix}
    \label{tab:Xmatrix}
    \vspace{-1.5\baselineskip}	
\end{table}

\vspace{0.5\baselineskip}	
\noindent
\textbf{Extraction Prioritization:}
In our study, \textit{Assignment} was prioritized highest, followed by \textit{File}, based on teachers' feedback, as it was the primary means of assessing student progress throughout the course before the final exam, and File was mainly used for uploading educational materials to support learning. Extraction was planned accordingly in the following sequence: \textit{submit assignment}, \textit{resubmit assignment}, \textit{set assignment grade}, \textit{view file}, \textit{view page}, \textit{view folder}, and \textit{set exam grade}. Ultimately, we implemented all the extractions so their order does not affect the final result.

\subsection{Log Extraction}

\noindent
\textbf{Implementation:}
We set up development and test environments to build and validate pipelines using Python that extracted relevant data objects and events, according to the extraction matrix, from Moodle’s relational database. We then transformed the extracted data into OCEL format. When transforming data into OCEL, we defined \textit{User} as a single object type, with \textit{Teacher} and \textit{Student} modeled as subclasses. This approach facilitated the discovery of process models at a general level while enabling drill-down analyses to distinguish teacher- and student-specific process behaviors. 

\vspace{0.5\baselineskip}	
\noindent
\textbf{Extraction:}
Data extraction was performed on a single course over one year. This timeframe provided a comprehensive dataset of interactions between students, teachers, and learning materials while ensuring manageability. By focusing on one course, we captured the full lifecycle of learning activities, including assignments, submissions, grading, and resource interactions.

\begin{figure}[t!]
    \adjustbox{center}{\includegraphics[width=1\linewidth]{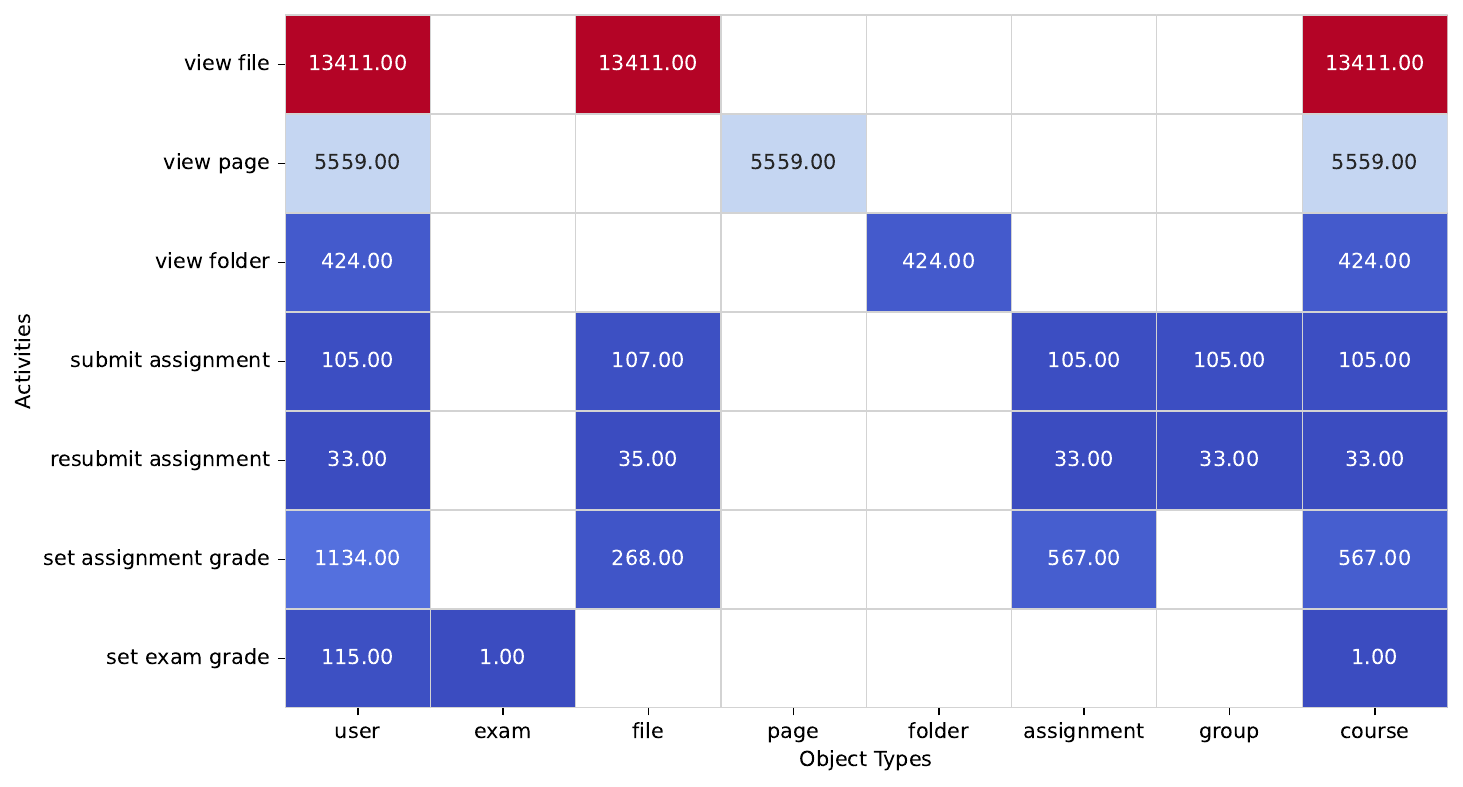}}
    \caption{The verification matrix}
    \label{fig:validationMatrix}
    \vspace{-1.5\baselineskip}	
\end{figure}

\vspace{0.5\baselineskip}	
\noindent
\textbf{Verification:}
We developed a script to systematically verify the accuracy of the extracted event data by mapping activities to object types. The resulting dataset contained 8 object types and 607 objects, linked through 7 event types and a total of 20,100 events, as shown in \figurename~\ref{fig:validationMatrix}. Since the data stems from a single course, all events are associated with this course, meaning the course column aggregates the total number of events.

\figurename~\ref{fig:validationMatrix} presents this structured validation process.
In this figure, the expected relationship between \textit{Group} and \textit{set assignment grade} is missing. This discrepancy arises because Moodle logs a separate \textit{set assignment grade} event for each student, even when a teacher grades a group assignment collectively. Automated verification helped identify such inconsistencies, highlighting areas for improvement in future log extractions by developers.
Furthermore, the relationship between \textit{Teacher} and \textit{Student} is not directly visible because the types of objects were recorded at the \textit{User} level. However, drill-down analysis can reveal these relationships as needed.

In this paper, we have presented only a subset of the case study conducted to demonstrate the application of our proposed methodology. Our full study included additional questions, object types, and events, during which we encountered further challenges in the extraction of OCEL, including the lack of recorded historical temporal data and instances of incomplete data capture.

An example of such challenges is the lack of recording historical temporal data in Moodle for module sections. Moodle only captures the current section for each module, so if a module like a File has been moved, the previous section will not be recorded in the extracted OCEL. We excluded presenting the section in this paper because it was outside the scope of the questions we selected for demonstration.

Another challenge related to incomplete data capture occurs when modules, like File, are imported from one course into another. In these cases, Moodle does not accurately log the importing user's details, resulting in the user information for these imported modules being unspecified. This incomplete logging limits the completeness of the extracted OCEL.

\subsection{Analysis Iteration}

We analyzed the extracted logs to answer the identified questions. The results were evaluated by the course responsible to ensure accuracy.

\begin{figure}[t!]
    \centering
    \adjustbox{center}{\includegraphics[width=1.1\linewidth]{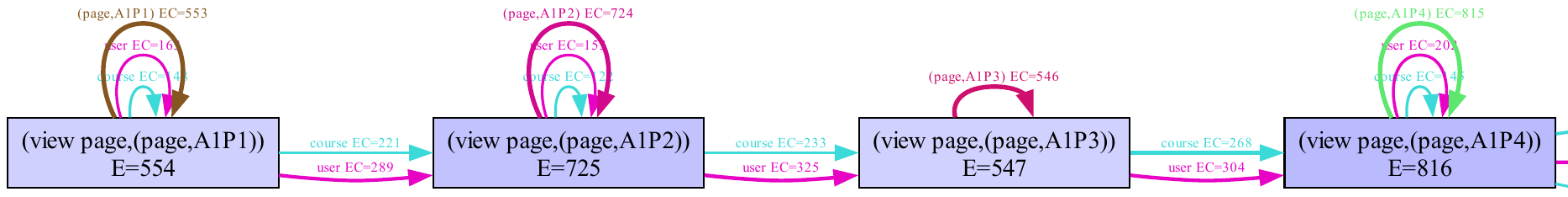}}
    \adjustbox{center}{\includegraphics[width=1.1\linewidth]{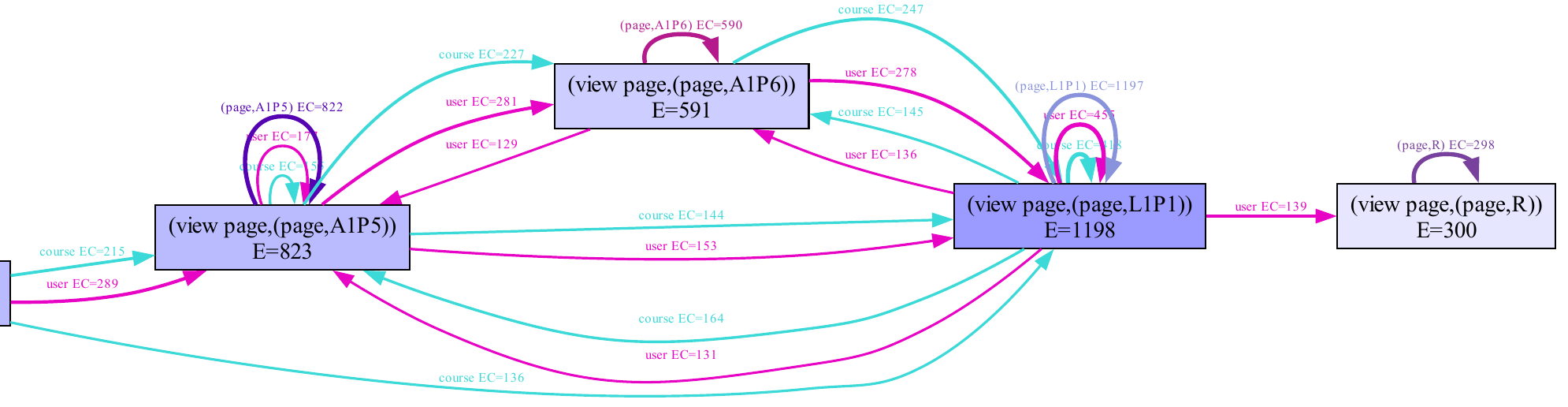}}
    \caption{Object-Centric Directly-Follows Graph (OC-DFG) illustrating how students visited some pages}
    \label{fig:verQ1}
    
\end{figure}

For Q1, \figurename~\ref{fig:verQ1} illustrates an Object-Centric Directly-Follows Graph (OC-DFG), discovered using the \textit{PM4Py} library. This graph was generated by filtering the extracted log, drilling down page objects into individual page names, and unfolding \textit{view page} events for each page using the \textit{processmining} library~\cite{khayatbashi2025olap}. The graph visualizes the sequential flow of actions during the course from the perspectives of the \textit{User} and \textit{Course} object types.

In this process, the page names for given codes are as follows: \textit{A1P1}: History of BPM, \textit{A1P2}: Business Process Models, \textit{A1P3}: Business Process Enactment, \textit{A1P4}: Syntax and Semantics, \textit{A1P5}: Control-flow Patterns, \textit{A1P6}: Business Process Complexity, and \textit{L1P1}: Introduction to Process Tree.
The model reveals that students sequentially accessed five pages: \textit{A1P1} to \textit{A1P5}. Additionally, a loop between \textit{A1P5}, \textit{A1P6}, and \textit{L1P1} suggests students frequently revisit these pages, indicating potential room for improvement.

\begin{figure}[t!]
    \centering
    \adjustbox{center}{\includegraphics[width=1.2\linewidth]{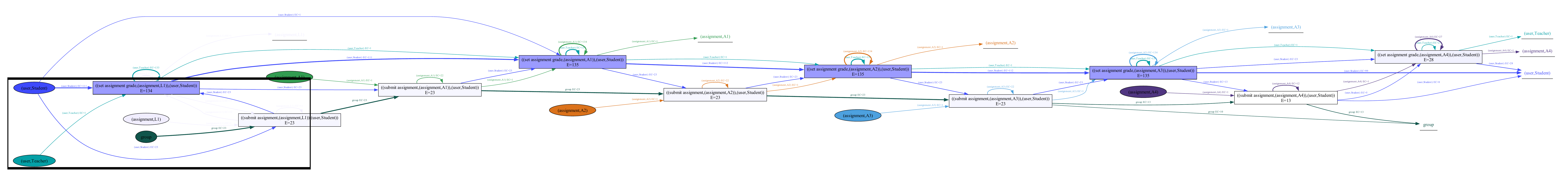}}
    \adjustbox{center}{\includegraphics[width=1.1\linewidth]{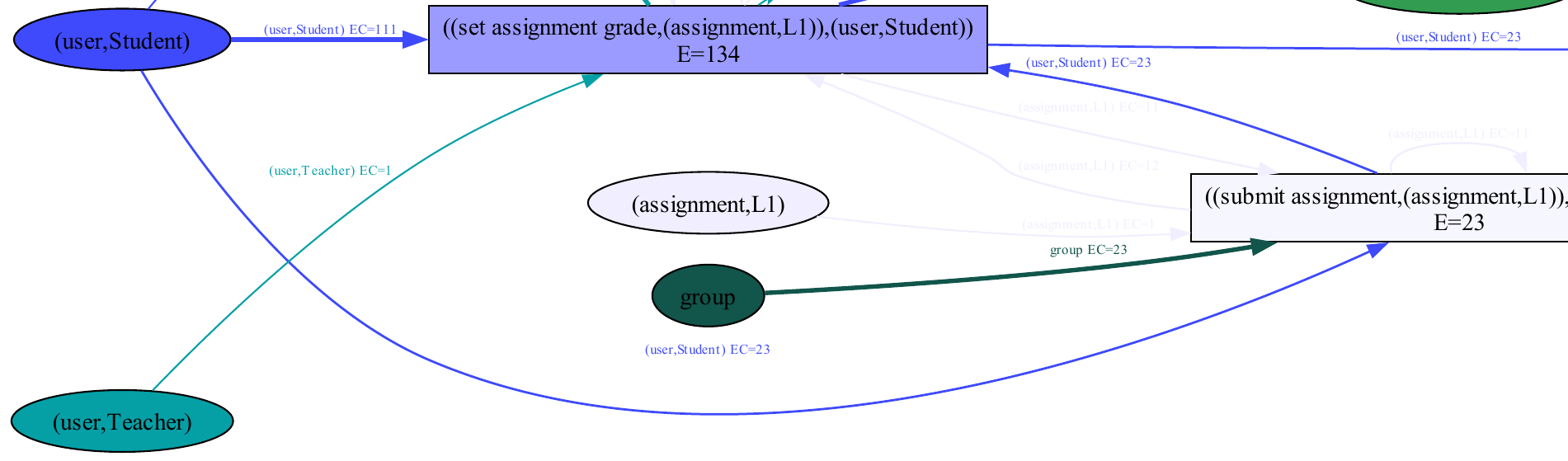}}
    \caption{OC-DFG illustrating the overall submission process for individual and group assignments in addition to a zoomed part}
    \label{fig:verQ2}
    \vspace{-1.5\baselineskip}	
\end{figure}

As an example of an analysis related to Q2, we refer to \figurename~\ref{fig:verQ2}, which presents an OC-DFG illustrating how individual and group assignments were submitted. The OC-DFG was discovered by filtering the log based on submission activities and events, drilling down user objects into user types (separating teachers and students) and assignments into their specific names (distinguishing different assignment object types). Additionally, the log was unfolded to include related events for these objects.

The upper part of the figure displays the entire process, and the lower part presents a zoomed-in version that crops the beginning of the process to enhance readability for further elaboration. The upper figure may not be easily readable (which is intentional, as we do not elaborate on details), but it effectively illustrates the sequential flow within the process. This sequence is influenced by the structured delivery of educational materials and the course setup.

In the lower part of the figure, we observe that 23 students submitted the assignment, as it was a group submission, where both the group and individual students were linked to the event. However, we also see that the grading process involved 134 students.
This discrepancy highlights the issue discussed in the verification section regarding the lack of a direct relationship between \textit{group} and \textit{set assignment grade}.

\vspace{-0.7\baselineskip}	
\subsection{Process Improvement \& Support}
Our analysis revealed that students frequently revisited pages related to control-flow perspectives, process complexity, and process trees, indicating possible challenges in comprehension or the sequence in which these topics are introduced. To address this, we propose shifting the process tree discussion to follow control-flow perspectives and introducing process complexity at the end. This adjustment may reduce unnecessary back-and-forth navigation, making the learning process more structured. Further investigation is needed to understand the underlying reasons for these recurring visits, providing insights that could help refine the course.

\vspace{-0.7\baselineskip}	
\section{Conclusion}\label{sec:Conclusion}
\vspace{-0.7\baselineskip}	
This paper introduced OCPM$^2$, an extension of the PM$^2$ methodology, designed to support Object-Centric Process Mining (OCPM) by providing a structured approach for Object-Centric Event Log (OCEL) extraction. Our methodology enhances existing process mining frameworks by addressing challenges associated with multi-object process modeling, ensuring compatibility with structured process analysis techniques.

By incorporating domain modeling, OCED extraction, and iterative process analysis phases, OCPM$^2$ provides a system-agnostic framework for object-centric process mining. The methodology reduces the complexity of extracting and analyzing multi-object event data, minimizing redundant log extraction efforts required for case-centric process mining.

To validate our approach, we applied OCPM$^2$ in an educational setting, extracting and analyzing OCELs from a learning management system and an administrative grading system, which cannot be shared due to privacy concerns. Our results demonstrated the practical usefulness of OCPM$^2$,  i.e., its ability to effectively reveal student navigation patterns and assignment submission processes.

The case study exposed limitations in system logs, particularly in capturing accurate object-to-activity relationships. A key issue was the missing link between \textit{Group} and \textit{set assignment grade}, causing Moodle to log separate grading events per student, even for collectively graded assignments. This discrepancy resulted in a mismatch between the number of students who submitted the assignment and those involved in the grading process. Automated verification identified these inconsistencies, helping developers extract correct event logs in future extractions.
Future research will focus on enhancing automated verification mechanisms, refining multi-dimensional process analysis operations, and expanding OCPM$^2$’s applicability across other domains, such as healthcare and financial services. We also plan to develop a tool to support OCED extraction based on OCPM$^2$. This tool will enable organizations to document and manage OCED extraction for different projects.

\bibliographystyle{plain}
\bibliography{main}

\begin{thebibliography}{10}

\bibitem{ocel2023specification}
A.~Berti, I.~Koren, J.~N. Adams, G.~Park, B.~Knopp, N.~Graves, M.~Rafiei, L.~Liß, L.~Tacke~Genannt Unterberg, Y.~Zhang, C.~Schwanen, M.~Pegoraro, and W.~M.~P. van~der Aalst.
\newblock {OCEL (Object-Centric Event Log) 2.0 Specification}, October 16 2023.
\newblock [Online].

\bibitem{2023SAPExtractionCaseStudy}
Alessandro Berti, Urszula Jessen, Gyunam Park, Majid Rafiei, and Wil Aalst.
\newblock Analyzing interconnected processes: using object-centric process mining to analyze procurement processes.
\newblock {\em International Journal of Data Science and Analytics}, pages 1--23, 07 2023.

\bibitem{2022SAPExtractionApproach}
Alessandro Berti, Gyunam Park, Majid Rafiei, and Wil Aalst.
\newblock {\em An Event Data Extraction Approach from SAP ERP for Process Mining}, pages 255--267.
\newblock 01 2022.

\bibitem{2009PDM-Bozkaya}
Melike Bozkaya, Joost Gabriels, and Jan~Martijn Van~der Werf.
\newblock Process diagnostics: A method based on process mining.
\newblock pages 22--27, 02 2009.

\bibitem{der2011discovery}
Van der Aalst and WMP~Process Mining.
\newblock Discovery, conformance and enhancement of business processes.
\newblock {\em Media; Springer: Berlin/Heidelberg, Germany}, 136, 2011.

\bibitem{esser2021multi}
Stefan Esser and Dirk Fahland.
\newblock Multi-dimensional event data in graph databases.
\newblock {\em Journal on Data Semantics}, 10(1):109--141, 2021.

\bibitem{fahland2024towards}
Dirk Fahland, Marco Montali, Julian Lebherz, Wil~MP van~der Aalst, Maarten van Asseldonk, Peter Blank, Lien Bosmans, Marcus Brenscheidt, Claudio di~Ciccio, Andrea Delgado, et~al.
\newblock Towards a simple and extensible standard for object-centric event data (oced)--core model, design space, and lessons learned.
\newblock {\em arXiv preprint arXiv:2410.14495}, 2024.

\bibitem{fowler2012patterns}
Martin Fowler.
\newblock {\em Patterns of enterprise application architecture}.
\newblock Addison-Wesley, 2012.

\bibitem{goossens2022enhancing}
Alexandre Goossens, Johannes De~Smedt, Jan Vanthienen, and Wil~MP van~der Aalst.
\newblock Enhancing data-awareness of object-centric event logs.
\newblock In {\em International Conference on Process Mining}, pages 18--30. Springer, 2022.

\bibitem{jalali2022object}
Amin Jalali.
\newblock Object type clustering using markov directly-follow multigraph in object-centric process mining.
\newblock {\em IEEE Access}, 10:126569--126579, 2022.

\bibitem{khayatbashi2023transforming}
Shahrzad Khayatbashi, Olaf Hartig, and Amin Jalali.
\newblock Transforming event knowledge graph to object-centric event logs: A comparative study for multi-dimensional process analysis.
\newblock In {\em International Conference on Conceptual Modeling}, pages 220--238. Springer, 2023.

\bibitem{khayatbashi2024transforming}
Shahrzad Khayatbashi, Olaf Hartig, and Amin Jalali.
\newblock Transforming object-centric event logs to temporal event knowledge graphs.
\newblock In {\em International Conference on Business Process Management}, pages 300--313. Springer, 2024.

\bibitem{khayatbashi2024advancing}
Shahrzad Khayatbashi, Najmeh Miri, and Amin Jalali.
\newblock Advancing object-centric process mining with multi-dimensional data operations.
\newblock {\em arXiv preprint arXiv:2412.00393}, 2024.

\bibitem{khayatbashi2025olap}
Shahrzad Khayatbashi, Najmeh Miri, and Amin Jalali.
\newblock {OLAP} operations for object-centric process mining.
\newblock In {\em Accpeted in {CAiSE} Forum 2025}. Springer, 2025.

\bibitem{khayabashi2025if}
Shahrzad Khayatbashi, Viktor Sjölind, Anders Granåker, and Amin Jalali.
\newblock {AI}-enhanced business process automation: A case study in the insurance domain using object-centric process mining, 2025.
\newblock {Accepted in BPMDS 2025}.

\bibitem{kimball2008data}
Ralph Kimball, Margy Ross, Warren Thornthwaite, Joy Mundy, and Bob Becker.
\newblock {\em The data warehouse lifecycle toolkit}.
\newblock John Wiley \& Sons, 2008.

\bibitem{2025uncovering_patterns}
Najmeh Miri and Amin Jalali.
\newblock Uncovering patterns in object-centric process mining: An approach using drill-down and roll-up techniques.
\newblock In Pari Delir~Haghighi, Michal Gregu{\v{s}}, Gabriele Kotsis, and Ismail Khalil, editors, {\em Information Integration and Web Intelligence}, pages 49--54, Cham, 2025. Springer Nature Switzerland.

\bibitem{2024moodle}
Moodle.
\newblock {\em Moodle Learning Management System}, 2024.
\newblock Version 4.4.

\bibitem{2025AnEduPMModel}
Eduardo Real and Edson Pimentel.
\newblock An educational process mining model on students’ paths data from virtual learning environments.
\newblock {\em Technology, Knowledge and Learning}, 2025.

\bibitem{2024ExploringObjectCentric}
Anukriti Tripathi, Aneesh, Yuvraj Shivam, Swetank Pandey, Aamod Vyas, and OP~Vyas.
\newblock Exploring object centric process mining with mimic iv: Unlocking insights in healthcare.
\newblock In {\em International Conference on Advanced Information Systems Engineering}, pages 360--372. Springer, 2024.

\bibitem{vanderAalst2022}
Wil M.~P. van~der Aalst.
\newblock {\em Process Mining: A 360 Degree Overview}, pages 3--34.
\newblock Springer International Publishing, Cham, 2022.

\bibitem{van2019object}
Wil~MP van~der Aalst.
\newblock Object-centric process mining: Dealing with divergence and convergence in event data.
\newblock In {\em Software Engineering and Formal Methods: 17th International Conference, SEFM 2019, Oslo, Norway, September 18--20, 2019, Proceedings 17}, pages 3--25. Springer, 2019.

\bibitem{2015PM2}
Maikel~L. van Eck, Xixi Lu, Sander J.~J. Leemans, and Wil M.~P. van~der Aalst.
\newblock {PM$^2$}: A process mining project methodology.
\newblock In Jelena Zdravkovic, Marite Kirikova, and Paul Johannesson, editors, {\em Advanced Information Systems Engineering}, pages 297--313, Cham, 2015. Springer International Publishing.

\bibitem{2023VKG}
Jing Xiong, Guohui Xiao, T.~Kalayci, Marco Montali, Zhenzhen Gu, and Diego Calvanese.
\newblock {\em A Virtual Knowledge Graph Based Approach for Object-Centric Event Logs Extraction}, pages 466--478.
\newblock 03 2023.

\end{thebibliography}

\end{document}